%% file: su3.tex
\documentclass[sigconf]{acmart}
\usepackage[utf8]{inputenc}
\usepackage{graphicx}
\usepackage{multirow}
\usepackage{multicol}
\usepackage{booktabs}
\usepackage{color}   
\usepackage{newfloat}
\usepackage{wrapfig}
\usepackage{url}
\usepackage{hyperref}
\usepackage{subfig}
\newcommand{\PIUMA}[1]{{{#1}}}
\title{Performance Optimization of SU3\_Bench on Xeon and Programmable Integrated Unified Memory Architecture}

\AtBeginDocument{%
  \providecommand\BibTeX{{%
    \normalfont B\kern-0.5em{\scshape i\kern-0.25em b}\kern-0.8em\TeX}}}

\begin{document}
\author{Jesmin Jahan Tithi}
\email{jesmin.jahan.tithi@intel.com}
\orcid{}
\author{Fabio Checconi}
\email{fabio.checconi@intel.com}
\author{Fabrizio Petrini}
\email{fabrizio.fetrini@intel.com}
\affiliation{%
  \institution{Parallel Computing Labs, Intel, CA, 95055 USA}
  \city{Santa Clara}
  \state{CA}
  \country{USA}
  \postcode{95055}
}

\author{Douglas Doerfler}
\affiliation{%
  \institution{Lawrence Berkeley National Laboratory}
  \city{Berkeley}
  \country{USA}}
\email{dwdoerf@lbl.gov}

\renewcommand{\shortauthors}{J.J. Tithi, et al.}

\input{abstract}

\maketitle
\input{introduction}
\input{Background}
\input{Performance_on_Xeon}
\input{Performance_on_PIUMA}
\input{Conclusion}
\bibliographystyle{IEEEtran}
\bibliography{su3}
\end{document}

%% file: abstract.tex
\begin{abstract}
 SU3\_Bench is a microbenchmark developed to explore performance portability across multiple programming models/methodologies using a simple, but nontrivial, mathematical kernel. This kernel has been derived from the MILC lattice quantum chromodynamics (LQCD) code. SU3\_Bench is bandwidth bound and generates regular compute and data access patterns. Therefore, on most traditional CPU and GPU-based systems, its performance is mainly determined by the achievable memory bandwidth. Although SU3\_Bench is a simple kernel, experience says its subtleties require a certain amount of tweaking to achieve peak performance for a given programming model and hardware, making performance portability challenging. In this paper, we share some of the challenges in obtaining the peak performance for SU3\_Bench on a state-of-the-art Intel Xeon machine, due to the nuances of variable definition, the nature of compiler-provided default constructors, how memory is accessed at object creation time, and the NUMA effects on the machine. We discuss how to tackle those challenges to improve SU3\_Bench's performance by \(2\times\) compared to the original OpenMP implementation available at Github. This provides a valuable lesson for other similar kernels.

Expanding on the performance portability aspects, we also show early results obtained porting SU3\_Bench to the new Intel Programmable Integrated Unified Memory Architecture (PIUMA), characterized by a more balanced flops-to-byte ratio. This paper shows that it is not the usual bandwidth or flops, rather the pipeline throughput, that determines SU3\_Bench's performance on PIUMA. Finally, we show how to improve performance on PIUMA and how that compares with the performance on Xeon, which has around one order of magnitude more flops-per-byte.\end{abstract}

%% file: introduction.tex
\section{Introduction}
SU3\_Bench \cite{SU3Git} is a microbenchmark developed at Lawrence Berkeley National Laboratory (LBNL) to explore performance portability across multiple programming models/methodologies using a simple, but nontrivial, mathematical kernel. This kernel has been derived from the MILC lattice quantum chromodynamics (LQCD) code~\cite{mimd2010milc}. The MILC Code is a body of high-performance  software written in C for doing SU(3) (special unitary group of degree 3) lattice gauge theory on high-performance computers, as well as single-processor workstations. The matrix-matrix and matrix-vector SU(3) operations are a fundamental building block of LQCD applications. Most LQCD applications use domain-specific implementations (libraries) written in machine-specific languages and/or intrinsics. Hence, performance portable methodologies are of interest. 

The SU3\_Bench microbenchmark calculates an SU(3) matrix-matrix multiply using complex floating-point arithmetic. The benchmark operates over a lattice of dimension = \(L^4\). The code is available at \url{https://gitlab.com/NERSC/nersc-proxies/su3_bench} under the LBNL modified BSD license.

SU3\_Bench is a bandwidth bound kernel. It is similar to the STREAM benchmark~\cite{STREAM,BabelStream} in the sense that it loads SU(3) matrices linearly from memory and stores the multiplication back to memory. Although SU3\_Bench has been optimized for Nvidia GPUs and AMD CPUs, this is the first work studying its optimization on Intel Xeon CPUs. In this paper, we discuss the challenges in obtaining close to peak performance for SU3\_Bench on state-of-the-art Intel Xeon systems and how the NUMA nature of the machine and the related data allocation policies may play a major role there. 

With the goal of exploring performance portability, we also show an early preview of SU3\_Bench’s performance on the new Intel Programmable Integrated Unified Memory Architecture (PIUMA). PIUMA's design is centered on programmability, performance portability, and scalability. It is a distributed global address space architecture optimized for irregular and sparse workloads and developed under the DARPA HIVE \cite{Hive2020piuma} program. PIUMA consists of a collection of multi-threaded cores, each equipped with a variety of offload engines, sharing a single global address space, supporting fine-grained memory and network accesses. PIUMA uses limited caching and small granularity memory accesses to deal efficiently with the memory behavior of sparse workloads. At the same time, PIUMA uses single-issue in-order pipelines with many threads to hide memory latency and avoid speculation. PIUMA supports in-network collectives, near-memory compute, and remote atomics in hardware. Unlike traditional Xeon Architecture, where the ``flops-to-byte'' ratio is often a number in the range of \(10\) to \(20\) or even higher, on PIUMA, this ratio is much smaller (\(1\) to \(3\) for example) and therefore, is not suitable for flop intensive workloads or kernels.

SU\_Bench's kernel has an arithmetic intensity of \(1.35\) for fp32 datatype (\(0.675\) for fp64 datatype) and appears to be a good match for PIUMA. In this paper, we show how to port SU3\_Bench to PIUMA and the types of optimizations needed to improve performance.

We make the following contributions:
\begin{itemize}
    \item We show a detailed analysis of SU3\_Bench's performance on Xeon architecture.
    \item We show how to overcome slowdown from NUMA effect on Xeon and get a \(2\times\) boost in performance on two sockets.
    \item We show how to port and optimize SU3\_Bench on PIUMA.
    \item We compare the performance of SU3\_Bench on Xeon vs. PIUMA.
\end{itemize}

\section{Related Work}
SU3\_Bench development to date has been primarily focused on evaluating various programming models for acceleration using general purpose GPU computing architectures~\cite{Doerfler2020P3HPCForum}. Although most of these programming models also support running on CPUs, tuning for a very wide SIMD architecture has not been done. This primarily involves remapping the data structures to better expose vectorization opportunities to the compiler. While this has been extensively studied for popular Lattice QCD domain specific library implementations, for example~\cite{10.5555/3050856}, these implementations also rely on architecture specific intrinsics.  Davis~\cite{Davis_WACCPD_2020} performed an extensive study of the OpenMP implementation using a wide variety of compilers, but only looking at target offload performance on GPUs. 

%% file: Background.tex
\section{Background}

\subsection{SU3\_Bench}
\label{SU3Background}
\begin{figure}[thpb]
	\centering
	\includegraphics[width=0.5\textwidth]{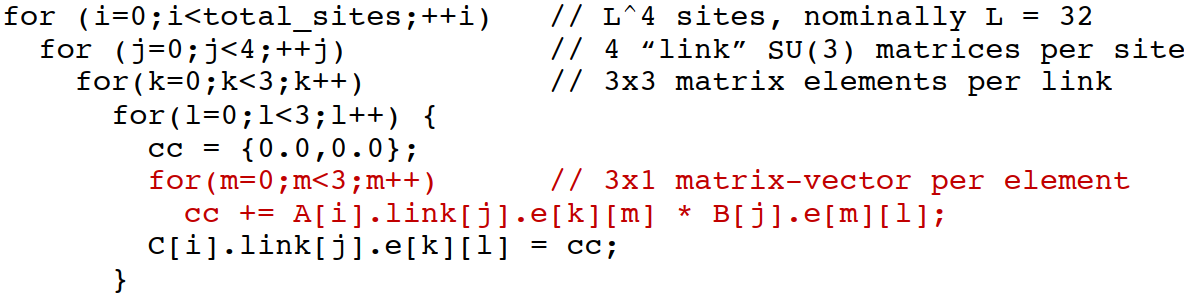}
	\caption{SU3\_Bench Core Kernel.}
	\label{fig:su3Kernel}
\end{figure}

In this section, we share details about the SU3\_Bench kernel. Figure \ref{fig:su3Kernel} shows how the kernel looks in its sequential form. The \(i\) loop iterates over \(L^4\) sites. Each site has four neighbors, or links, and the \(j\) loop iterates over them. For each such link, it computes a 3x3x3 general matrix multiplication (GEMM) among the complex numbers (real, img) representing the links using \(k\), \(l\), and \(m\) loops. There are 3x3 matrix elements per link and the innermost \(m\) loop is essentially doing a dot product, multiplying a row of \(A[i]\)'s element matrix with a column of \(B[j]\)'s link matrix.
This kernel has \(\Theta(n)\) compute and \( \Theta(n)\) memory load and store operations and hence is memory-bound on all (or most) modern architectures. 

\begin{figure}[thpb]
	\centering
	\includegraphics[width=0.5\textwidth]{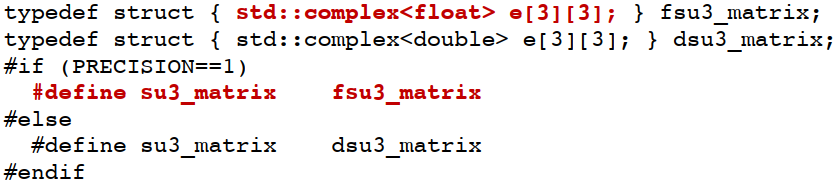}
	\caption{Su3\_Matrix Data Structure.}
	\label{fig:Matrix}
\end{figure}

\begin{figure}[thpb]
\vspace{-10pt}
	\centering
	\includegraphics[width=0.5\textwidth]{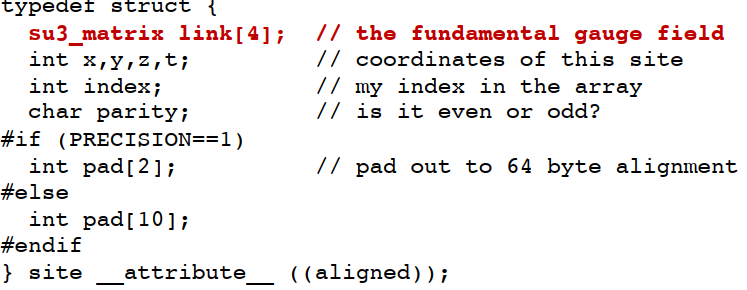}
	\caption{Site Data Structure.}
	\label{fig:siteMatrix}
\end{figure}

We first consider two key data structures used in the kernel. Figures \ref{fig:Matrix} and \ref{fig:siteMatrix} show the definitions for the \(su3\_matrix\) and \(Site\) data structures. An \(su3\_matrix\) is a 3x3 matrix of complex type: for fp32 values it requires \(72\) bytes (\(144\) bytes for fp64) of storage. The \(Site\) definition is based on MILC’s \cite{mimd2010milc} lattice.h, but reduced to the bare minimum fields. It contains four links of type \(su3\_matrix\), coordinates of this site, index in the large site array, and whether the parity is even or odd. It has some padding to make it a multiple of \(64\) and requires \(320\) bytes for fp32 and \(640 \) bytes for fp64 values.

Next, we consider the sizes of the input and output arrays and their implications for performance. When \(L=32\) in Figure \ref{fig:su3Kernel}, the size of site array \(A\) would be \(320\)\,MiB for fp32 (\(32^4 \times 320 = 320\)) and \(640\)\,MiB for fp64. Similarly, the output array \(C\) would be of the same size as \(A\). There is no expected cache re-use for \(A\) and \(C\) outside of the GEMMs. The size of \(B\) is \(288\) bytes for fp32 data, \(576\) bytes for fp64, and this size is constant. \(B\) could stay in the cache and can be reused. Note that, in most state-of-the-art Xeon machines this data (\(A\), \(B\), and \(C\)) will not fit in the L3 cache, which is often of size \(40\) \,MiB or less. A and C data would usually need to be streamed from/to memory.

According to Figure \ref{fig:su3Kernel}, the number of floating point operations for each site is \(4 \times (3 \times 3 \times 3)) \times (4 ~\mathrm{mul} + 4 ~\mathrm{add})\) = \(4 \times (108 ~\mathrm{mul} + 108 ~\mathrm{add})\) = \(4 \times 216 ~\mathrm{ops} = 4 \times 216 = 864\). The data size for each of A[i] and C[i] is 320 Bytes for fp32 (single precision) and 640 Bytes for fp64 (double precision) values. Therefore, the arithmetic intensity (AI) is \(864/(320\times2)\) = 1.35 for fp32 data and 0.675 for fp64 data. This calculation ignores reading from B.

Since performance portability is a goal for SU3\_Bench, it can be used as a tool to explore the performance implications and programmability on any existing or future architectures. In this paper, we analyze the performance portability of SU3\_Bench on the state-of-the-art Xeon and the novel PIUMA architecture that we discuss next.

\begin{figure*}[h]
	\centering
	\includegraphics[width=0.9\textwidth]{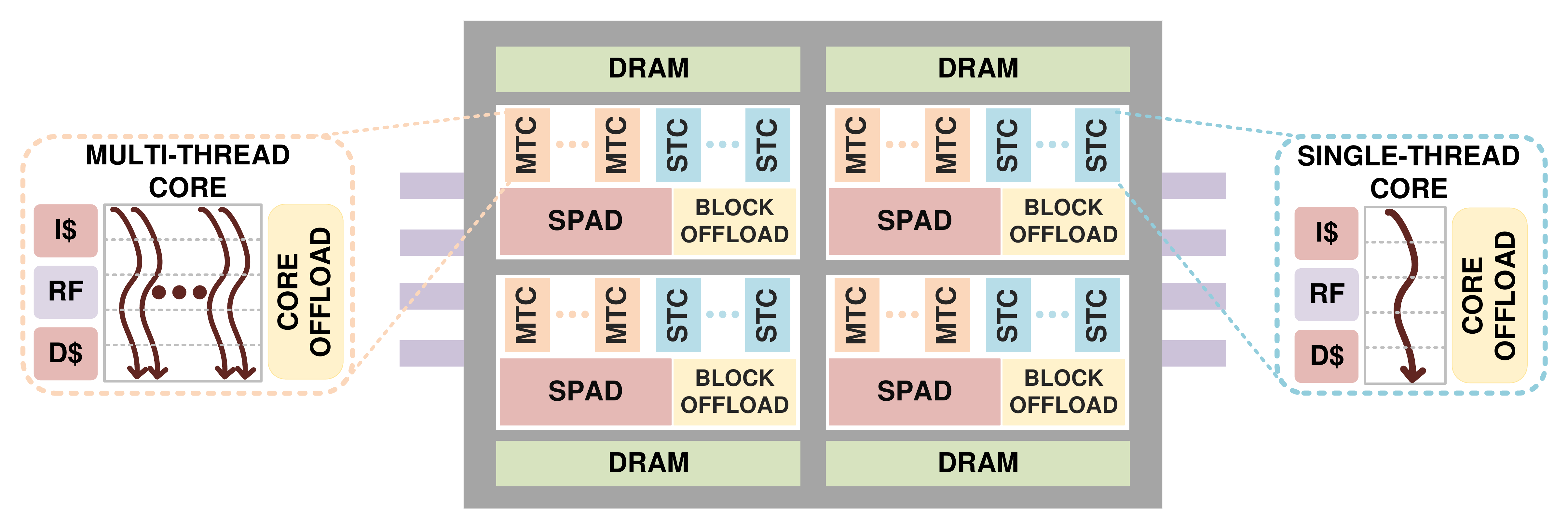}
	\caption{High-level diagram of PIUMA architecture (collected from \cite{Hive2020piuma}).}
	\label{fig:PIUMA_arch_overview}
\end{figure*}

\subsection{PIUMA}
\label{PIUMAArch}

In this section, we give a detailed description of the PIUMA Architecture. This information has been reproduced from \cite{Hive2020piuma}. The PIUMA architecture \cite{PIUMABlog,HIVEWiki,Hive2020piuma} consists of a collection of highly multi-threaded cores (MTC) and single-threaded cores (STC) as shown in Figure~\ref{fig:PIUMA_arch_overview}. The MTCs are round-robin multi-threaded to address the lack of instruction-level parallelism in most sparse workloads and incorporate latency hiding through thread-level parallelism instead of aggressive out-of-order or speculative execution models. At any moment, each thread can only have one in-flight instruction, which considerably simplifies the core design for better energy efficiency \cite{Hive2020piuma}. However, this also can be a limiting factor when an operation (say fused multiply-add) requires several data to be loaded before the operation can be executed.

While the MTCs are the data parallel engines in the PIUMA architecture, 
the STCs are used for single thread performance sensitive tasks, such as memory and thread management (e.g., from the operating system). These
are in-order stall-on-use cores that are able to exploit some
instruction and memory-level parallelism, while avoiding the
high power consumption of aggressive out-of-order pipelines.
Both core types implement the same custom RISC instruction
set. \cite{Hive2020piuma}

\PIUMA{
All MTCs and STCs in PIUMA have a local instruction cache (I\$), 
data cache (D\$), and register file (RF). PIUMA supports 
selective data caching through the use of a unique bit in 
the address space. Coherency between data caches is included 
to support a variety of programming models. For scalability, 
caches are not coherent across the whole system. It is the 
responsibility of the programmer to avoid modifying shared 
data that is cached, or to flush caches if required for correctness. MTCs
and STCs are grouped into blocks, each of which has a large
local scratchpad (SPAD) as low latency storage. Programmers
are responsible for selecting which memory accesses to cache
(e.g., local stack), which to put on SPAD (e.g., frequently reused
data structures or the result of a DMA gather operation) and
which not to store locally. There are no prefetchers to avoid
useless data fetches and to limit power consumption. Instead,
the offload engines can be used to efficiently
fetch large chunks of useful data. Each core's offload region in Figure~\ref{fig:PIUMA_arch_overview} 
contains a direct memory access (DMA) engine that executes gather, 
scatter, copy, initialization, reduction, and broadcast operations. 
The DMA engine supports executing atomic operations at the remote 
destinations. \cite{Hive2020piuma}}

\PIUMA{
PIUMA implements a hardware distributed global address
space (DGAS), which enables each core to uniformly access
memory across the full system with
one address space. Besides avoiding the overhead of setting
up communication for remote accesses, a DGAS model also greatly
simplifies programming, because there is no implementation
difference between accessing local and remote memory. This is also great in terms of performance portability. Address 
translation tables contain programmable rules to
translate application memory addresses to physical locations,
to arrange the address space to the need of the application
(e.g., address interleaved, block partitioned, etc.).
The memory controllers (one per block) can
support native 8-byte accesses, while supporting standard cache
line accesses as well. \cite{Hive2020piuma}}

\PIUMA{
PIUMA has a high-radix, low diameter HyperX topology network. Each link in the PIUMA network is optimized for small user messages (sub 16-byte) to avoid the software and hardware inefficiencies of message aggregation and large buffers prevalent in traditional systems. This architecture utilizes a small-granularity interface with concurrent transactions to achieve similar aggregate bandwidths as traditional implementations. In PIUMA, network bandwidth exceeds local DRAM bandwidth, which is different from conventional architectures that assume higher local traffic than remote traffic. The high-radix, low-diameter network design along with the DGAS implementation reduces the overall access latency from the perspective of the instruction pipelines. \cite{Hive2020piuma}
}

\PIUMA{Since PIUMA targets sparse workloads, the supported\footnote{The quoted architectural flops-per-byte is experimental and is subject to change.} flop-per-byte is low compared to the traditional architectures. The SU3\_Bench kernel has a flop-per-byte of 0.675 for fp64 number. This is lower than the flop-per-byte achievable on PIUMA and therefore, SU3\_Bench for fp64 should not be bounded by the compute on PIUMA.}

%% file: Performance_on_Xeon.tex
\section{Performance on Xeon}

In this section, we discuss the challenges in obtaining peak performance on a state-of-the-art Intel Xeon system, and how the subtleties of variable definition, the nature of compiler provided default constructor, and the NUMA architecture of the machine affect the performance of SU3\_Bench.

The SU3\_Bench code is written in standard C and C++. A common main driver routine is used for all programming model implementations, with implementations specific to each model self-contained in respective C++ include files. Su3\_Bench has different versions available for different platforms (CPU, GPUs, FPGAs) and different programming environments (CUDA, Intel dpcpp, hip, hipsycl, OpenACC, OpenCL, OpenMP, sycl). Since we are interested in performance on Xeon, we used the simplest and most popular: the OpenMP version.

The OpenMP implementation of SU3\_Bench has four versions. 
\begin{itemize}
\item {\bf Version \(0\)} uses ``\#pragma omp target teams distribute num\_teams (num\_teams) thread\_limit (threads\_per\_team)'' on the \(i\) loop and collapses the \(j\), \(k\), and \(l\) loops. Version \(0\) is our baseline. 

\item {\bf Version \(1\)} uses ``\#pragma omp target teams num\_teams (num\_teams) thread\_limit (threads\_per\_team)'' as a parallel block and then the total sites are distributed manually across teams. Again the \(j\), \(k\), and \(l\) loops are collapsed into one.

\item {\bf Version \(2\)} uses work items per thread approach with work\\\_items = total\_sites * THREADS\_PER\_SITE following a typical CUDA programming style. Then the "\#pragma omp target teams distribute parallel for num\_teams (num\_teams) thread\_limit (threads\_per\_team)'' is used on the work item loop. The \(i\), \(j\), \(k\), \(l\) indices are computed from the work item ids. This essentially eliminates \(j\), \(k\), and \(l\) loops which are now manually collapsed into the \(i\) loop. 

\item {\bf Version 3} has ``\#pragma omp target teams distribute parallel for collapse(4) num\_teams (num\_teams) thread\_limit (threads\_per\_team)'' on the \(i\) loop. It is instructing to collapse all four loops. 
\end{itemize}

We started by testing all four versions in the code on the following Intel Xeon Platform and compiler:
\begin{itemize}
\item CPU Name: Intel(R) Xeon(R) Platinum 8280 CPU @ 2.70GHz, CPU MHz (1800) turbo on (CLX 8280)
\item Memory: 196 \,GiB (DRAM), 39.4 \,MiB (L3), 1 \,MiB (L2), 32 \,KiB (L1)
\item Memory bandwidth: 105 GB/s (1 socket) and 210 GB/s (2 sockets) 
\item Stream bandwidth on two sockets: 200 GB/s
\item 2 Sockets: 56 cores, 56 threads
\item Parallelism: Shared-memory thread parallel
\item Compiler in Xeon: Intel(R) 64 Version 2020
\end{itemize}

The KMP\_AFFINITY was set to ``compact,1,0,granularity.'' Using KMP\_AFFINITY=compact,1,0 binds the \({n+1}^{th}\) OpenMP thread to a thread context as close as possible to the \(n^{th}\) thread, but on a different core. Once each core has been assigned one OpenMP thread, the subsequent OpenMP threads are assigned to the available cores in the same order, but they are assigned on different thread contexts. The number of teams was set to \(56\) and threads per team were set to 1. There was an additional outermost loop that runs the core kernel (Figure~\ref{fig:su3Kernel}) a given number of iterations (I) and warmup (W) times to provide stable performance data.

\begin{table*}[hptb]
  \centering
  \caption{Roof\/line Analysis on Xeon}

    \begin{tabular}{rrrrrrrrr}
      & \multicolumn{8}{c}{\textbf{with 2x SIMD units}} \\
\cmidrule{2-9}        & \multicolumn{1}{c|}{SIMD=8} & \multicolumn{1}{c|}{SIMD=7} & \multicolumn{1}{c|}{SIMD=6} & \multicolumn{1}{c|}{SIMD=5} & \multicolumn{1}{c|}{SIMD=4} & \multicolumn{1}{c|}{SIMD=3} & \multicolumn{1}{c|}{SIMD=2} & \multicolumn{1}{c}{SIMD=1} \\
\midrule
     Single socket & 141.8 & 141.8 & 141.8 & 141.8 & 141.8 & 141.8 & 141.8 & 141.8 \\
     Single core & 86.4 & 75.6 & 64.8 & 54.0 & 43.2 & 32.4 & 21.6 & 10.8 \\
        & \multicolumn{8}{c}{\textbf{with 1x SIMD units or 2x units w/no FMA}} \\
        \midrule
\cmidrule{2-9}        & \multicolumn{1}{c|}{SIMD=8} & \multicolumn{1}{c|}{SIMD=7} & \multicolumn{1}{c|}{SIMD=6} & \multicolumn{1}{c|}{SIMD=5} & \multicolumn{1}{c|}{SIMD=4} & \multicolumn{1}{c|}{SIMD=3} & \multicolumn{1}{c|}{SIMD=2} & \multicolumn{1}{c}{SIMD=1} \\
     Single socket & 141.8 & 141.8 & 141.8 & 141.8 & 141.8 & 141.8 & 141.8 & 141.8 \\
     Single core & 43.2 & 37.8 & 32.4 & 27.0 & 21.6 & 16.2 & 10.8 & 5.4 \\
        & \multicolumn{8}{c}{\textbf{with 1x SIMD units \& no FMA}} \\
\cmidrule{2-9}        & \multicolumn{1}{c|}{SIMD=8} & \multicolumn{1}{c|}{SIMD=7} & \multicolumn{1}{c|}{SIMD=6} & \multicolumn{1}{c|}{SIMD=5} & \multicolumn{1}{c|}{SIMD=4} & \multicolumn{1}{c|}{SIMD=3} & \multicolumn{1}{c|}{SIMD=2} & \multicolumn{1}{c}{SIMD=1} \\
\midrule
     Single socket & 141.8 & 141.8 & 141.8 & 141.8 & 141.8 & 141.8 & 141.8 & 75.6 \\
     Single core & 21.6 & 18.9 & 16.2 & 13.5 & 10.8 & 8.1 & 5.4 & 2.7 \\
    \bottomrule
    \end{tabular}%
  \label{tab:roofline}%
\end{table*}%
At this point, we can revisit the roof\/line analysis for this platform in a little more detail.
Each core runs at 2.7\,GHz clock speed, has 2 SIMD units with 8 lanes per SIMD unit and each of those lanes can execute two flops. Therefore, the maximum GFLOPS per second (GF/s) on this core is = 86.4 = 2.7 \,GHz \(\times\) 2 SIMD units \(\times\) 8 lanes/SIMD unit \(\times\) 2 FLOPs/lane. On single socket with 28 cores, maximum GF/s = 2420.1 = 86.4 \,GF/s/core \(\times\) 28 cores. The maximum bandwidth per single socket is = 105.0 GB/s = 2.933 GHz \(\times\) 8 B/channel \(\times\) 6 channels. Therefore, the flops per byte ratio of this architecture is 17.1 = 2420.1/105.0. The arithmetic intensity of the SU3\_Bench kernel is 1.35 for fp32 data type and therefore, SU3\_Bench will be bandwidth bound on Xeon.

Table~\ref{tab:roofline} shows the theoretical roof\/line analysis of SU3\_Bench on this Xeon platform.
Because of the way data structures have been implemented in SU3\_Bench, it is difficult for compilers to identify vectorization opportunities for the inner loop, especially for CPUs with wide SIMD units. It is the goal of SU3\_Bench developers to address this in a future version of the benchmark.
Assuming only 1 lane per each of the 2 SIMD units are used, the peak performance for a single-core is 10.8\,GF/s and for a single socket is 141.8\,GF/s. The peak performance with 1 SIMD unit and 1 lane per unit utilization is 5.4\,GF/s for a core and 141.8\,GF/s for a socket. Lastly, the peak performance with 1 SIMD unit and 1 lane per unit without any FMA utilization is 2.7\,GF/s for a core and 75.6\,GF/s for a socket. Details of this analysis can be found in Table~\ref{tab:roofline}.

Table~\ref{tab:Baseline} shows the initial performance of the four versions of SU3\_Bench on two sockets CLX 8280. Overall, the performance on two sockets is close to half of what we expected from the roof\/line analysis. Version 2 performs the best when there is only one warm-up and one iteration (-I 1 -W 1), or 100 iterations and 1 warm-up (-I 100 -W 1). However, if we run 200 iterations and 1 warm-up (-I 200 -W 1), Version 1 performs the best. Notice that if we increase the number of iterations, that improves performance. The question is why increasing number of iterations would improve performance. We first consider a fixed cost overhead for thread setup and cache warmup, but we note that the value of \(L\) in this case is \(32\), so none of the arrays other than B would fit in cache completely, and data should be streamed from memory. However, the maximum bandwidth obtained was only \(70\) GB/s which is even lower than what one socket should offer.

\begin{table*}[htbp]
\vspace{-8pt}
  \centering
  \caption{Baseline Performance Data}
  {
   
    \begin{tabular}{ccccccc}
    \toprule
        & \multicolumn{2}{c}{\textbf{-I200-W1}} & \multicolumn{2}{c}{\textbf{-I100-W1}} & \multicolumn{2}{c}{\textbf{-I1-W1}} \\
    \midrule
    \textbf{VERSIONS} & \textbf{GFLOPS} & \textbf{GBYTES} & \textbf{GFLOPS} & \textbf{GBYTES} & \textbf{GFLOPS} & \textbf{GBYTES} \\
    \midrule
    \textbf{version 0} & 89.783 & 66.506 & 70.870 & 52.500 & 66.760 & 49.450 \\
    \textbf{version 1} & 94.820 & 70.237 & 72.940 & 54.030 & 56.570 & 41.910 \\
    \textbf{version 2} & 92.542 & 68.550 & 73.042 & 54.110 & 69.960 & 51.830 \\
    \textbf{version 3} & 33.675 & 24.945 & 33.590 & 24.880 & 33.900 & 25.140 \\
    \bottomrule
    \end{tabular}%

    }
  \label{tab:Baseline}%
\end{table*}%

The next likely candidate is NUMA imbalance. To verify how it affects performance we rerun the above experiment, this time adding numactl -C 0-55 -N 0,1 -m 0,1 to the command line while running the program. We have already instructed OpenMP runtime to use a compact affinity. This instructs Linux to bind OS processes 0 -- 55 to cores 0 -- 55 and use both nodes 0 and 1 for the compute resources and memory allocation. Table~\ref{tab:BaselineNuma} shows the performance data.

\begin{table*}[htbp]
\vspace{-5pt}
  \centering
  \caption{Baseline Performance Data With Numa Control}
    {

    \begin{tabular}{ccccccc}
    \toprule
    \multicolumn{7}{c}{\textbf{With NUMACTL -C 0-55 -N 0,1 -m 0,1 }} \\
    \midrule
        & \multicolumn{2}{c}{\textbf{-I200-W1}} & \multicolumn{2}{c}{\textbf{-I100-W1}} & \multicolumn{2}{c}{\textbf{-I1-W1}} \\
    \midrule
    \textbf{VERSIONS} & \textbf{GFLOPS} & \textbf{GBYTES} & \textbf{GFLOPS} & \textbf{GBYTES} & \textbf{GFLOPS} & \textbf{GBYTES} \\
    \textbf{version 0} & 72.309 & 53.562 & 69.768 & {51.68} & 68.885 & 51.030 \\
    \textbf{version 1} & 74.157 & 54.931 & 72.877 & 53.983 & 71.460 & 52.930 \\
    \textbf{version 2} & 73.140 & 54.178 & 72.692 & 53.846 & 71.389 & 52.887 \\
    \textbf{version 3} & 33.721 & 24.978 & 33.754 & 25.119 & 33.911 & 25.119 \\
    \bottomrule
    \end{tabular}%
    }
  \label{tab:BaselineNuma}%
\end{table*}%

Controlling NUMA allocation improves performance for versions 0, 1, and 2 for a single iteration (-I 1 -W 1). However, the obtained bandwidth is still \(53\) GB/s --- only half of the single-socket streaming bandwidth. Also notice that the performance degraded for \(100\) or \(200\) iterations compared to not using numactl, shown in Table~\ref{tab:Baseline}. As a result, the maximum bandwidth obtained is reduced to only \(55\) GB/s and it did not change much by increasing the number of iterations. As a potential candidate to explain the dependency on the number of iterations we turned our attention to page migration. The Linux memory management subsystem, unless directed otherwise, supports migrating pages between NUMA nodes to improve the locality of the accesses. Use of numactl can prevent automatic page migrations. Manual NUMA tuning of applications (i.e., use of numactl) overrides automatic NUMA balancing, disabling periodic unmapping of memory, NUMA faults, migration, and automatic NUMA placement of an application \cite{NUMA}.

Before digging deeper into the NUMA issues, we wanted to have a simple baseline version of the code --- one with the simple and most popular OpenMP pragma which is also easy to understand and debug. Despite having four variants, the SU3\_Bench OpenMP implementation is missing a very basic OpenMP loop structure of "\#pragma omp parallel for" on the \(i\) loop. Therefore, we added the simplest possible implementation of SU3\_Bench with "\#pragma omp parallel for" in the \(i\) loop. We call it VersionX. The performance of this version is shown in Table~\ref{tab:versionX}. VersionX is faster than all prior OpenMP versions of SU3\_Bench. Nevertheless, the best bandwidth obtained was less than the stream bandwidth of one socket. Additionally, we noticed the same trend with and without numactl as we have seen for Versions 0, 1, and 2. That is: without numactl and with an increased number of iterations, performance improves but with numactl, the performance remains close to the single iteration. 
\begin{table*}[htbp]
\vspace{-10pt}
  \centering
  \caption{Performance of VersionX}
  {

    \begin{tabular}{ccccccc}
    \toprule
 
        & \multicolumn{2}{c}{\textbf{-I200-W1}} & \multicolumn{2}{c}{\textbf{-I100-W1}} & \multicolumn{2}{c}{\textbf{-I1-W1}} \\
    \midrule
    \textbf{Runs} & \textbf{GFLOPS} & \textbf{GBYTES} & \textbf{GFLOPS} & \textbf{GBYTES} & \textbf{GFLOPS} & \textbf{GBYTES} \\
    \textbf{without NUMACTL} & 100.843 & 74.698 & 76.442 & 56.620 & 73.300 & 54.300 \\
    \textbf{with NUMACTL} & 75.026 & 55.575 & 74.930 & 55.500 & 73.495 & 54.440 \\
    \bottomrule
    \end{tabular}%

    }
  \label{tab:versionX}%
\end{table*}%

To get more information on the runs, we used the Intel Vtune tool to analyze the performance of VersionX. We found that although the application reported bandwidth (effective bandwidth) was only \(\approx 82\) GB/s, the actual bandwidth noted by Vtune is around \(182\) GB/s --- almost double of what the useful bandwidth was. This 182 GB/s is close to 210 GB/s, the maximum offered by the system. It also maxed out the system's UPI bandwidth indicating substantial cross-socket traffic.
\begin{figure*}[t!]
	\centering
	\includegraphics[width=0.7\textwidth]{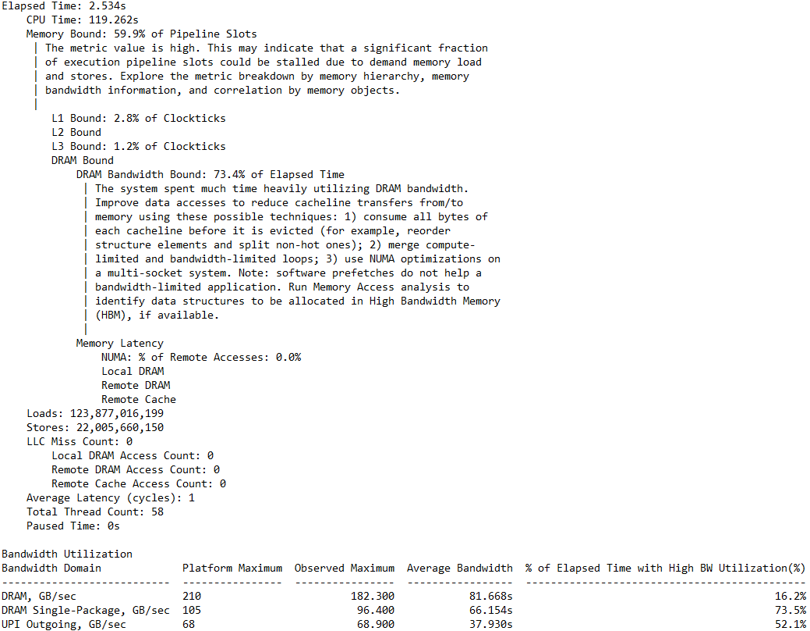}
	\caption{Vtune Analysis of VersionX of SU3\_Bench without NUMACTL and 200 iterations.}
	\label{fig:vtune}
\end{figure*}

We developed a custom library to print the allocated memory in each socket and Table~\ref{tab:pageMigration} shows that with an increased number of iterations, the allocated memories in the two sockets are changing. It is happening because, when the numactl is off, the Linux is migrating pages from one node to the other to move the pages closer to the node where the thread is. The more number of iterations it gets, the better it becomes. That is why as we increase the number of iterations, the better the performance becomes. However, if numactl is used, the automatic page migration stops and as a result, the performance remains the same other than reducing the thread launching overheads.

\begin{table}[htbp]
  \centering
  \caption{Total page sizes at different sockets}
    \begin{tabular}{cccc}
    \toprule
    \textbf{Total Mem (B)} & \textbf{Socket 0} & \textbf{Socket 1} & \textbf{Iterations} \\
    \midrule

    6.71E+08 & 6.78E+08 & 3.70E+06 & 1 \\
    6.71E+08 & 6.79E+08 & 5.21E+06 & 2 \\
    6.71E+08 & 3.44E+08 & 3.42E+08 & 200 \\

    \bottomrule
    \end{tabular}%
  \label{tab:pageMigration}%
\end{table}%

\begin{figure}[htbp]
  \centering
    \begin{tabular}{l}
    std::vector<site> a(total\_sites); \\
    std::vector<su3\_matrix> b(4); \\
    std::vector<site> c(total\_sites); \\
    \end{tabular}%
     \caption{Variable Declaration in SU3\_Bench.}
  \label{fig:var}%
\end{figure}%
The above results also indicate that during data allocation, the data is mainly placed in the first socket. This made sense since the data initialization routine in SU3\_Bench is a sequential function. The thread that executes the main method is likely to be in the first socket. The main thread allocates the data and based on the "first touch policy" the data is allocated at socket 0. 

Therefore, we parallelized the data initialization routine following the exact same pattern as the data is accessed during compute hoping that each thread would touch the memory in the same way as it does during compute, and hence data will be allocated on the appropriate sockets. However, that did not improve runtime at all. To investigate what went wrong, we used our custom library to print the amount of data allocated on two sockets before and after the initialization and it turns out that the data got allocated and touched on socket 0 during the declaration of the variables itself, i.e., after the lines shown in Figure~\ref{fig:var} and therefore, the parallel initialization does not have much impact. 
\begin{table*}[htbp]
  \centering
  \caption{Amount of Data Allocated on Both Sockets}
   \resizebox{1\textwidth}{!}{
    \begin{tabular}{rlr}
    \toprule
    \multicolumn{1}{|p{16.44em}|}{Before any initialization                                                                  } & \multicolumn{1}{p{16.19em}|}{After vector creation                                                   } & \multicolumn{1}{p{15.565em}|}{After data initialization                                                                  } \\
    \midrule
    \multicolumn{1}{|p{16.44em}|}{Node 0 MemTotal:       99300960 kB                                                         } & \multicolumn{1}{p{16.19em}|}{Node 0 MemTotal:       99300960 kB                                                         } & \multicolumn{1}{p{15.565em}|}{Node 0 MemTotal:       99300960 kB                                                         } \\
    \multicolumn{1}{|p{16.44em}|}{Node 0 MemFree:        92942092 kB                                                         } & \multicolumn{1}{p{16.19em}|}{Node 0 MemFree:        92277780 kB                                                         } & \multicolumn{1}{p{15.565em}|}{Node 0 MemFree:        92277780 kB                                                         } \\
    \multicolumn{1}{|p{16.44em}|}{Node 0 MemUsed:         6358868 kB                                                         } & \multicolumn{1}{p{16.19em}|}{Node 0 MemUsed:         7023180 kB                                                         } & \multicolumn{1}{p{15.565em}|}{Node 0 MemUsed:         7023180 kB                                                         } \\
    \multicolumn{1}{|p{16.44em}|}{\textbf{Node 0 Active:            13260 kB                                                         }} & \multicolumn{1}{p{16.19em}|}{\textbf{Node 0 Active:           674936 kB                                                         }} & \multicolumn{1}{p{15.565em}|}{\textbf{Node 0 Active:           674936 kB                                                         }} \\
    \multicolumn{1}{|p{16.44em}|}{Node 1 MemTotal:       100663296 kB                                                        } & \multicolumn{1}{p{16.19em}|}{Node 1 MemTotal:       100663296 kB                                                        } & \multicolumn{1}{p{15.565em}|}{Node 1 MemTotal:       100663296 kB                                                        } \\
    \multicolumn{1}{|p{16.44em}|}{Node 1 MemFree:        97572436 kB                                                         } & \multicolumn{1}{p{16.19em}|}{Node 1 MemFree:        97568284 kB                                                         } & \multicolumn{1}{p{15.565em}|}{Node 1 MemFree:        97568284 kB                                                         } \\
    \multicolumn{1}{|p{16.44em}|}{Node 1 MemUsed:         3090860 kB                                                         } & \multicolumn{1}{p{16.19em}|}{Node 1 MemUsed:         3095012 kB                                                         } & \multicolumn{1}{p{15.565em}|}{Node 1 MemUsed:         3095012 kB                                                         } \\
    \multicolumn{1}{|p{16.44em}|}{\textbf{Node 1 Active:             2112 kB }} & \multicolumn{1}{p{16.19em}|}{\textbf{Node 1 Active:             5724 kB }} & \multicolumn{1}{p{15.565em}|}{\textbf{Node 1 Active:             5724 kB                                                         }} \\
    \midrule
        & std::vector\(< site>\) a(total\_sites); & \multicolumn{1}{l}{make\_lattice\_range(a.data(), istart, istop, ldim, Complx\{1.0,0.0\});} \\
        & std::vector\(<su3\_matrix>\) b(4); & \multicolumn{1}{l}{init\_link(b.data(), Complx\{1.0/3.0,0.0\});} \\
        & std::vector\(<site>\) c(total\_sites); &  \\
        \bottomrule
    \end{tabular}%
    }
  \label{tab:socketData}%
\end{table*}%

It contradicts our expectation that these variables would get first touched (and hence allocated) during the initialization routine. Instead, data got allocated during the declaration as shown in Table~\ref{tab:socketData}. It explains why we did not get performance improvement from the parallel initialization. We thought it must be something to do with how the memory is first touched which in turn means how the constructors for \(site\) and \(su3\_matrix\) are being called. It turns out that these structures do not have any user-defined constructors and therefore must be using/calling compiler provided default constructors which by default initialize everything with zeros. These default constructors must be touching the memory during the declaration (Figure~\ref{fig:var}) and as a result, all the data was getting allocated on socket 0 and consequently, nothing was changing during the parallel initialization.

To solve this issue, we added empty constructors in the structures that do "nothing" (i.e., do not touch anything). However, since now these constructors exist, the compiler-provided default constructor will not be called anymore and thus, those memory locations will not be touched during declaration. This time, when we use the parallel initialization, we see that both sockets have roughly an equal amount of memory allocated on them. This solves our problem with imbalanced data on sockets and inefficient bandwidth usage due to page migrations by Linux. This change improves performance by \(2.6\times\) and we are now able to obtain \(143.46\)\,GB/s and \(193.54\)\,GF/s on 2 sockets (56 cores) of CLX8280. 

\begin{figure*}[tpbh]
    \begin{tabular}{cc}
        \includegraphics[width=0.5\textwidth]{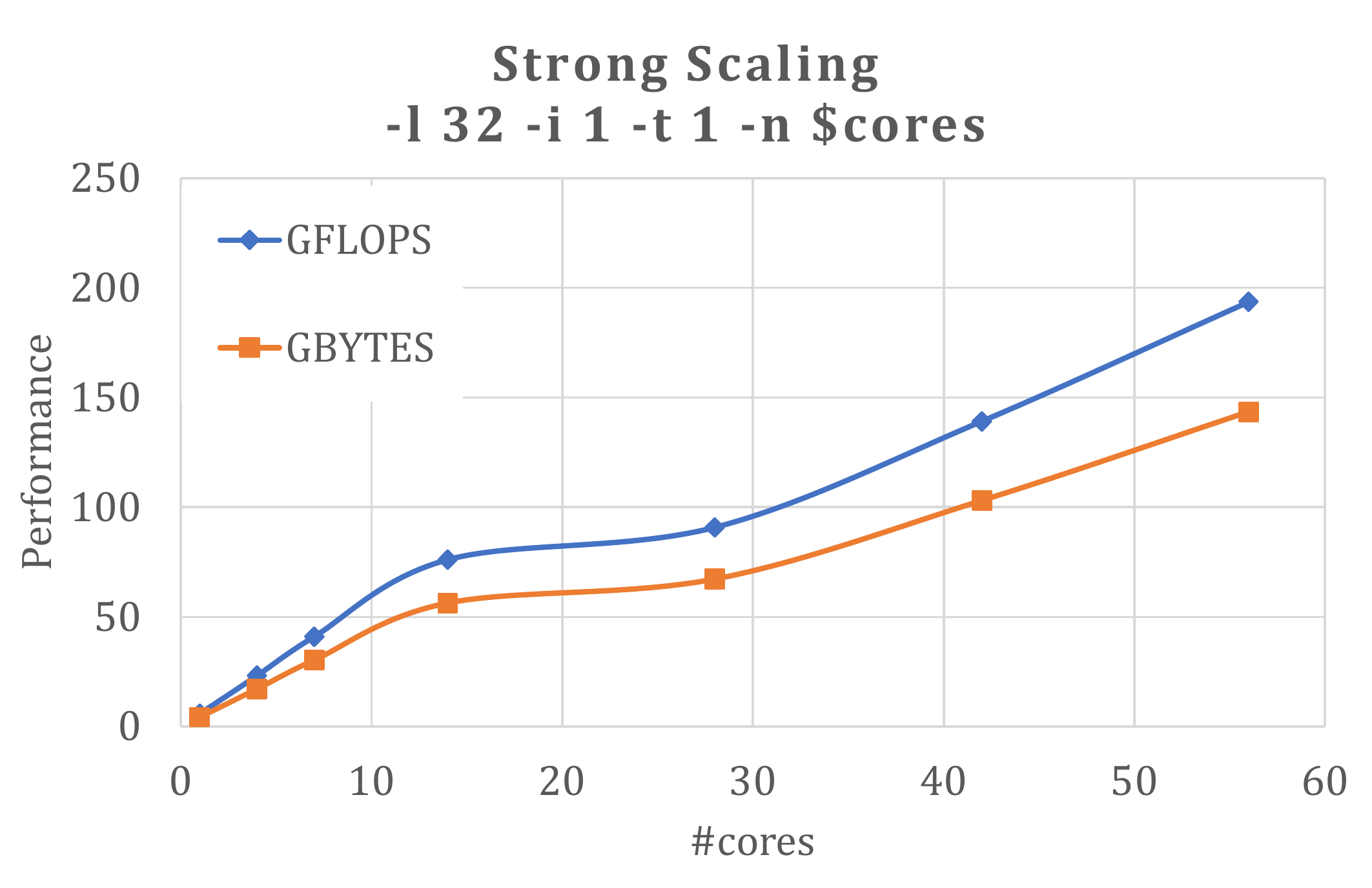} &
        \includegraphics[width=0.5\textwidth]{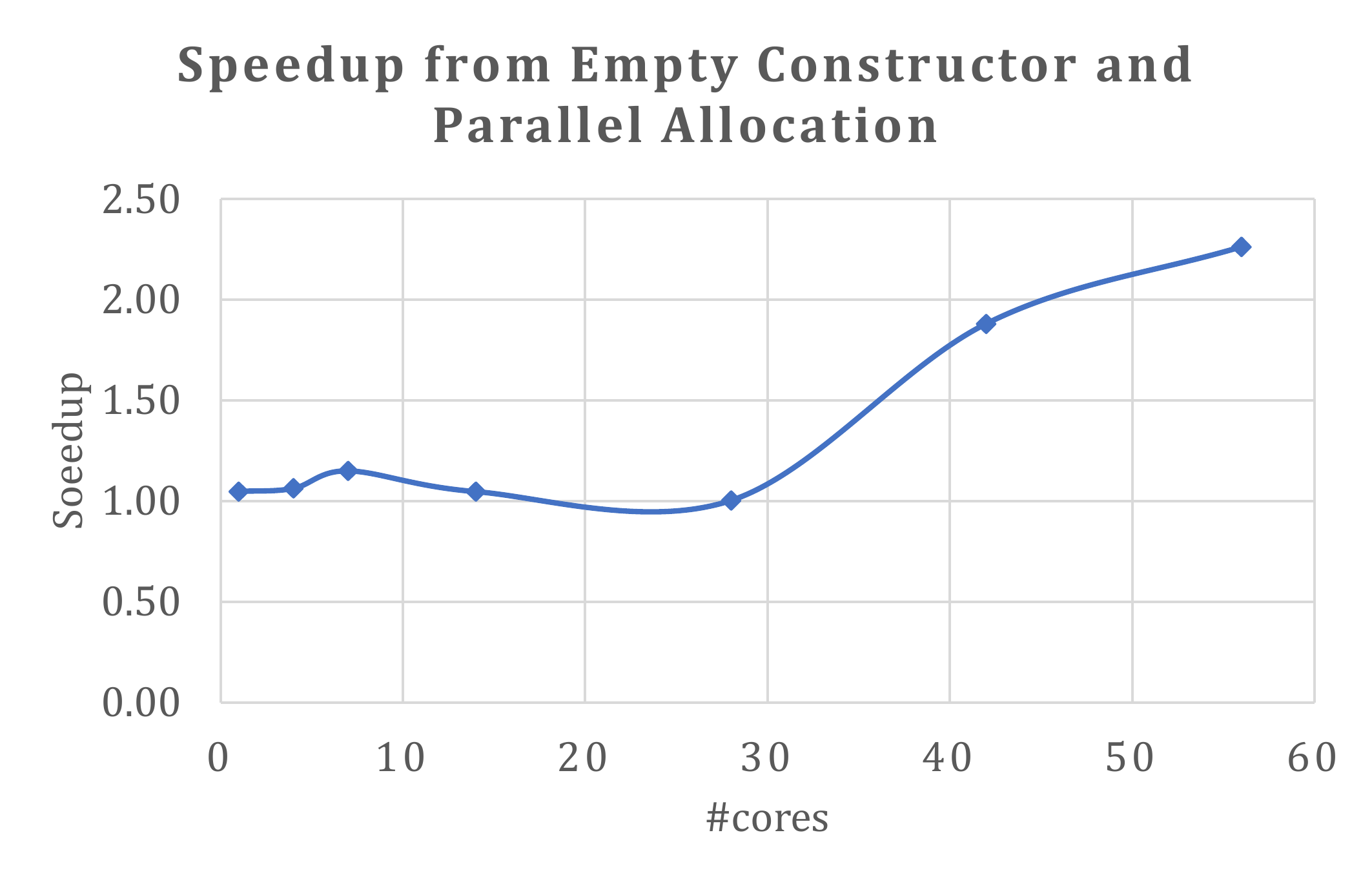}
    \end{tabular}
    \caption{VersionX: Strong Scaling}
    \label{fig:stongscaling}
\end{figure*}
Note that at this point, \(A\) and \(C\) vectors are allocated on both sockets based on the first touch policy. However, \(B\) is first touched (and allocated) sequentially. Furthermore, \(B\) is accessed in column-major order (non-unit stride). Hence, it might be more efficient to copy it to the local cache of each thread in a transposed manner. We adopted this optimization. 

Figure~\ref{fig:stongscaling} (left side) shows strong scaling of VersionX of SU3\_Bench. We see a linear increase in performance from \(1\) to \(14\) cores, minimal performance improvement from \(14\) to \(28\) cores since it had already saturated the single socket bandwidth with \(14\) cores), then a linear performance increase from \(29\) to \(56\) cores, albeit, with a smaller slope than 1 to 14 cores. This strong scaling trend is exactly as we expected for a non-NUMA software (i.e., a software whose performance is not sensitive to NUMA issue). Figure~\ref{fig:stongscaling} (right side) shows till 28 cores (i.e., within a single NUMA domain), the performance for both with and without the empty constructor is similar. However, beyond one socket, the version with an empty constructor is over \(2\times\) faster than the original one. The total speedup obtained using \(56\) cores (compared to one core run) is \(32.4\times\). 

\begin{figure*}[tpbh]
    \begin{tabular}{cc}
        \includegraphics[width=0.5\textwidth]{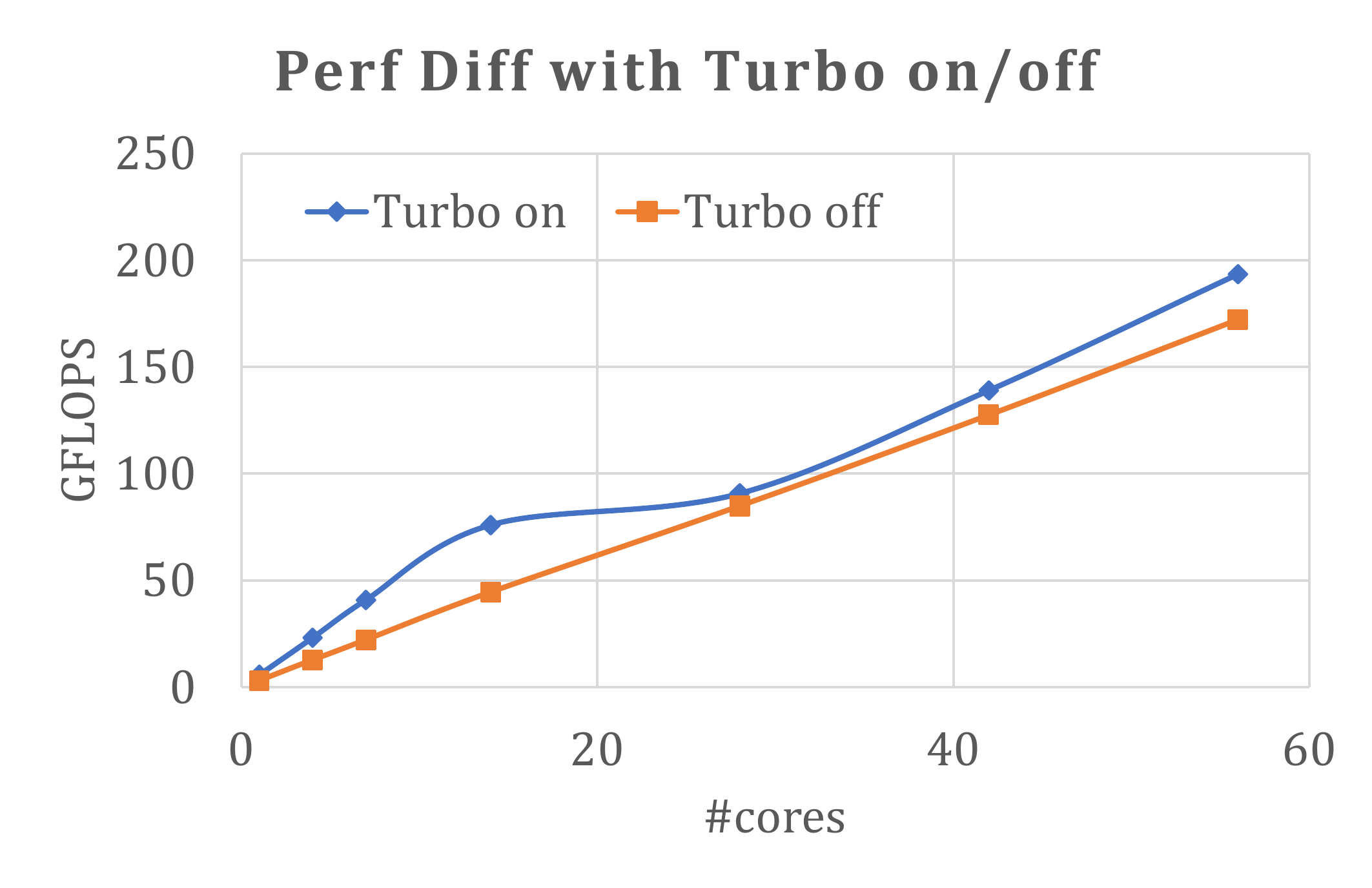} &
        \includegraphics[width=0.5\textwidth]{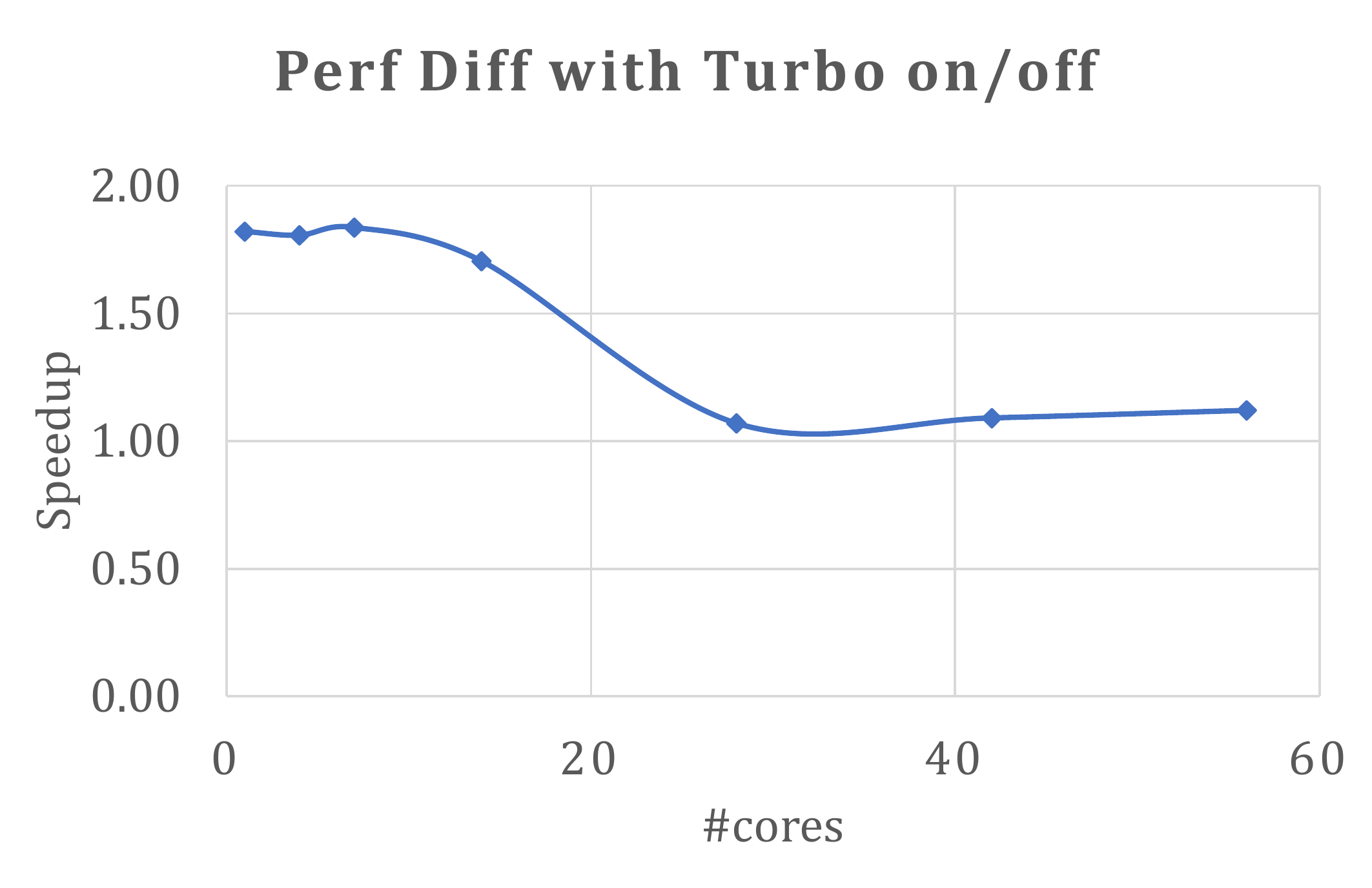}
    \end{tabular}
    \caption{VersionX: Benefits of using Turbo Boost.}
    \label{fig:turbo}
\end{figure*}

We also tested the impact of turbo boost on the performance as shown in Figure~\ref{fig:turbo}. Turbo boost can improve performance up to \(1.8\times\), especially at low core counts and the benefit depletes as the number of cores increases and reaches a saddle point at \(28\) cores (full socket) and then when we add cores from the second socket, we start to see some performance benefit. However, the curve for turbo "off" appears to be more regular and stable than that of turbo "on". This matches what we know about turbo boost in general.

Note that for a bandwidth-bound kernel the expected maximum speedup from 56 cores is around \(26\times\) assuming one core can drive up to 8\,GB/s. We believe, this implementation of SU3\_Bench obtaining \(32.4\times\) speedup indicates some form of data reuse and potential inefficiency of the single core run which drives \(4.34\),GB/s bandwidth.

At this point, although the performance improvement on two sockets is satisfactory and relatively close to the peak, the single-core performance (5.85\,GF/s) appeared to be indicative of the case where only 1 SIMD unit is used or the case where 2 SIMD units are used without any FMA. This motivated us to use explicit general matrix multiplication (GEMM) and FMA instructions instead of depending on the compiler to do so. We tried replacing the \(k, l, m\) loops with an explicit 3x3x3 GEMM routine that manually called the FMA instructions after unrolling all products. This improved performance by \(1.6\times\) -- \(1.8\times\) up to \(8\) cores. However, at \(28\) cores, the performance improvement is only \(11\)\%, and at 56 cores, it is down to \(0\)\%. We observe a significant variance in performance from run to run at 56 cores (with and without turbo) for this version. Additionally, we had to use the compiler flag -xCORE-AVX512 and without this flag, the performance goes down below the original version. This suggests that doing the manual GEMM allowed it to better utilize the AVX vectorization. The single-core performance of this version reaches the peak performance of 2 SIMD units (10.4\, GF/s). The total speedup with 56 cores, in this case, is \(18.4\times\) with respect to its one core run. Full SIMD utilization is a known issue in the Lattice QCD community, and specialized libraries have been developed for Intel Xeon CPUs with very high bandwidth memory subsystems, such as the Intel Xeon Phi~\cite{10.5555/3050856}. Those types of optimizations are outside the scope of this paper. 

We checked how L1, L2, and L3 prefetches impact the performance of SU3\_Bench on Xeon. Turning off L1 prefetch has almost no impact on performance, turning off L2 prefetch drops performance by \(5\)\% at 56 cores. However, if we turn off all three levels of prefetch, performance drops by over \(2\times\). So, we can conclude that prefetching at L3 makes the difference.

\begin{figure*}[tpbh]
    \begin{tabular}{cc}
        \includegraphics[width=0.5\textwidth]{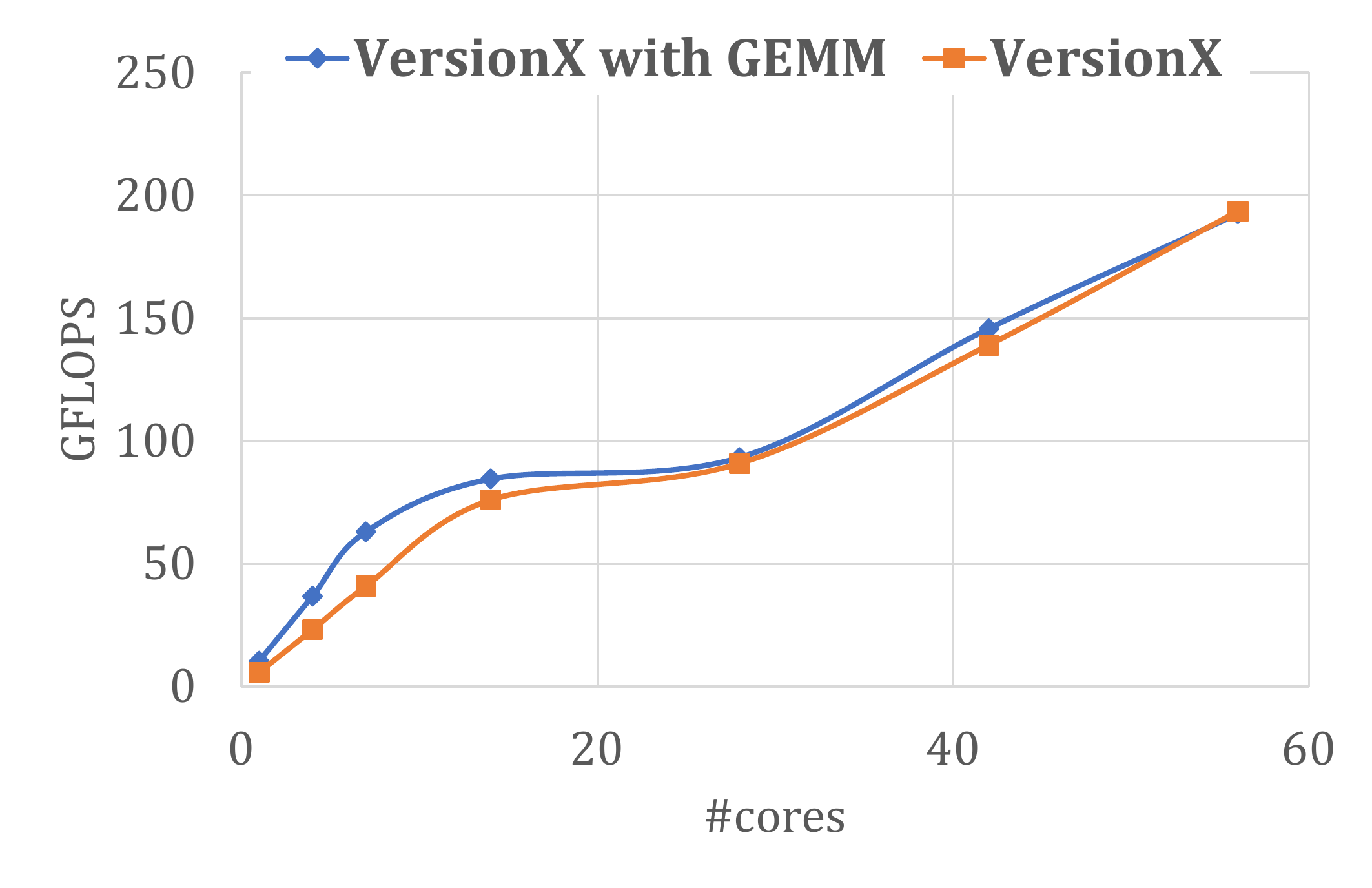} &
        \includegraphics[width=0.5\textwidth]{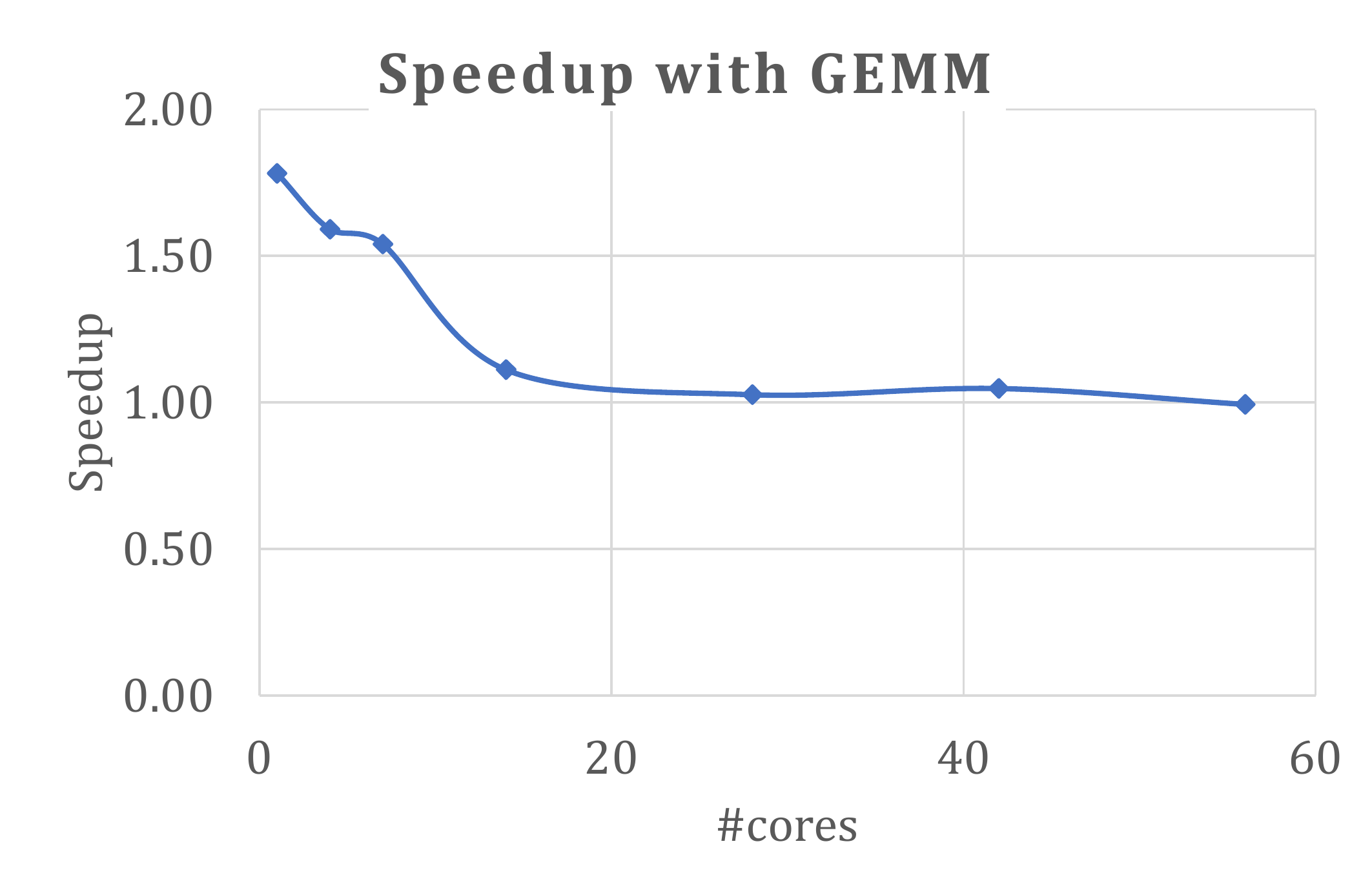}
    \end{tabular}
    \caption{VersionX: Benefits of using GEMM instead of Dot product.}
     \label{fig:mm}
\end{figure*}

Notice that even at this point we did not reach the STREAM bandwidth of the system. Obtaining STREAM~\cite{STREAM,BabelStream} bandwidth would require us to be able to use streaming write instructions. Since the \(site\) data structure has 320\,bytes size and we are only writing 288\,bytes of it, the compiler is unable to use streaming write because there are gaps in between each write.

To summarize the lessons learned in this section, we can say that making sure that the data is close to compute is crucial for performance of SU3\_Bench. Not providing a default constructor could create an unforeseeable problem by touching memory at creation. Using an empty constructor and then using a parallel initialization routine solves the issue. Simple OpenMP pragma provides decent performance. Using turbo boost improves performance. Replacing dot product with explicit GEMM also improves performance. However, if we are running on the full system, GEMM has no positive impact on the performance.

%% file: Performance_on_PIUMA.tex
\section{SU3\_Bench on PIUMA}
In this section, we show some early preview of SU3\_Bench's performance on the new PIUMA architecture. PIUMA has been designed with programmability in mind. It supports C and many features of C++. It has its own OpenMP style programming extensions to exploit both Single Program Multiple Data (SPMD) and task-based parallelization schemes.

\subsection{Initial Porting Effort}
It's fairly easy to port an OpenMP or pthread style shared-memory code to PIUMA since any memory location can be read/written by simple load/store operations (thanks to DGAS). We ported VersionX of SU3\_Bench to PIUMA without the latter optimizations first (GEMM and \(B\) transpose optimization). We added those optimizations later. We made the following changes in the code:
\begin{itemize}
    \item Changed the memory allocation calls to match the memory allocation library of PIUMA which are specialized for DGAS allocations.
    \item Changed the OpenMP parallelization to explicit PIUMA style SPMD parallelization by dividing the number of sites equally among the threads.
\end{itemize}

We also changed the main program to include PIUMA library header files and the Makefile to support PIUMA runtime. We used the STCs for memory allocations and the MTCs to initialize and execute the main kernel in parallel. PIUMA allows a program to selectively cache data by turning on a given bit in the address of that data. However, data is not automatically prefetched to cache as happens in Xeon. We initially chose to cache both \(A\) and \(B\). 

Since PIUMA supports a variety of memory allocation options, the porting process involves deciding how to allocate various data structures in memory. There is no first-touch policy involved here. Data gets allocated based on the instruction provided. By default, PIUMA uses an interleaved/striped memory allocation. For a program running on \(M\)~blocks with \(M\)~memory controllers, any memory allocation will be striped in round-robin chunks across those memory controllers. This helps ensure even access distribution across the memory controllers and reduces queueing latency and conflicts for randomly accessed data. For our initial porting effort, we allocated everything in the main memory, striped across memory controllers. This does not appear to be the best allocation policy for SU3\_Bench, which has strictly sequential access. The best option would be to allocate data close to the memory controller where the accessing threads are (based on our experience on Xeon). However, we show here that since PIUMA is a DGAS system and the access latency for local or remote data is not too different and since the network bandwidth is not lower than the local memory bandwidth, we are able to reach close to the peak bandwidth despite the striped allocation policy.

We run this code on PIUMA's cycle-accurate simulator, which is over \(10,000\times\) slower compared to running on actual hardware, therefore severely restricting our capability to simulate large problems and more iterations. To reduce simulation time, we removed warmups on PIUMA while running the experiments. 
\subsection{Performance on PIUMA}
When we run this program with L=16, we obtained 2.14\,GF/s and an average dram bandwidth utilization of 74\%, with an IPC (instructions Per Cycle) of 3.7. 
On the same input, CLX8280 obtains 5.519\,GF/s and 4.385 GB/s. Therefore, with this version, the PIUMA core is \(2.57\times\) slower than the Xeon core. The performance gap is more than the ratio\footnote{The number might vary in the actual hardware.} of bandwidths (1.56).

To understand what to expect in terms of performance on PIUMA, we show the roof\/line analysis in the following section. 
\subsection{Roof\/line Analysis on PIUMA}
The compute-bound performance of SU3\_Bench on PIUMA is 8\, GF/s if all multiply-add is done using FMAs. If no FMA is used, the peak flop achievable is 4\, GF/s.

The bandwidth-bound performance on PIUMA is 4.32\,GF/s.

Apart from the above two, there is a third aspect of PIUMA that may dictate the performance of the SU3\_Bench kernel and that is the rate of instruction issues per cycle. Since PIUMA has simple single-issue in-order scalar cores, these cores are not able to issue one FMA per cycle, due to the need to load or store the data each FMA accesses on the same scalar pipeline. For example, if we revisit the SU3\_Bench kernel shown in Figure~\ref{fig:su3Kernel}, we see in the innermost dot product loop that for each element of \(C\) we need 12 FMAs and at least 12 loads for \(A\) and \(B\) in total and 2 stores for \(C\). In other words, to produce 24 FLOPs, at least 12 Loads, 2 stores, and 12 FMAs need to be executed. 
This leads to 
3.6\,GF/s per PIUMA core. Therefore, according to roof\/line analysis, we can only expect to get the minimum of the above three as the maximum performance, i.e., 3.6\,GF/s from single PIUMA core.

Unlike Xeon, were we use only FLOPs or Bandwidth for roof\/line analysis, on PIUMA we also had to consider the instruction execution rate which was the limiting factor. Our initial implementation on PIUMA did not reach that peak.

\subsection{Optimization on PIUMA}
We inspected the generated code and realized that it had a few extra instructions to handle register spilling and the compiler was unable to generate FMA instructions for all the updates. We realized that the 3x3x3 complex GEMM might be too big for the simple cores of PIUMA, causing frequent register spills. 
 
Therefore, a blocked GEMM of size 2x3 of \(A\) times 3x3 of \(B\) followed by 1x3 of \(A\) times 3x3 of \(B\) could be a better approach. The blocked multiplication of \(A\)[2x3] by \(B\)[3x3] requires 12 loads from \(A\), 18 loads from \(B\), and does 12 stores to \(C\). Also, it requires \(2 \times 3 \times 3\) complex multiplications, or 72 FMAs. A blocked multiplication of \(A\)[1x3] by \(B\)[3x3] requires 6 loads from \(A\), 18 loads from \(B\), and does 6 stores to \(C\) and requires \(1 \times 3 \times 3\) complex multiplications, or 36 FMAs. Overall, the upper bound of FLOPs when limited by instruction issue rate in this case is \(2 \times (72 + 36) / (12 + 18 + 12 + 72 + 6 + 18 + 6 + 36) \) = 1.2\,GF/s per pipeline and 4.8\,GF/s per PIUMA core. Replacing dot product with blocked GEMM improves the roof\/line bound as well.

We tried this blocked GEMM and with L=32, and 4 iterations, with fp64 (double precision) data, we get 3.72\,GF/s, and a bandwidth of 5.1\,GB/s. For the same input, one core of the CLX 8280 obtains a total of 5.94\,GF/s and 8.8\,GB/s. With the blocked GEMM, one PIUMA core is \(1.57\times\) slower than the CLX core which is an improvement over the last version and now this difference can be justified by the bandwidth difference among the two platforms. When we run the same problem on 8 PIUMA cores, we get 24.56\,GF/s, and 34.38\,GB/s. CLX 8280 can get a total 35.79\,GF/s and a total bandwidth of 53.03\,GB/s. Therefore, 8 PIUMA cores are \(1.42\times\) slower than 8 CLX cores. Again, this performance is consistent with the gap in bandwidth between the two. However, the speedup from 8 cores on PIUMA is \(6.61\times\) and for CLX, it is \(6.04\times\). On 16 cores of PIUMA, we get 43.88\,GF/s and a total bandwidth of 61.42\,GB/s. On 16 CLX cores, SU3\_Bench gets 43.731\,GF/s and 64.787\,GB/s. Therefore, 16 cores of PIUMA is on par with 16 CLX cores. At 32 cores, PIUMA is \(1.48\times\) faster than 32 cores of CLX. This happens mainly due to the difference between the effective bandwidth that each of these platforms can obtain. PIUMA cores get 108\,GB/s whereas Xeon gets 73.8\,GB/s.

\begin{figure}[ht!]
	\centering
	\includegraphics[width=0.5\textwidth]{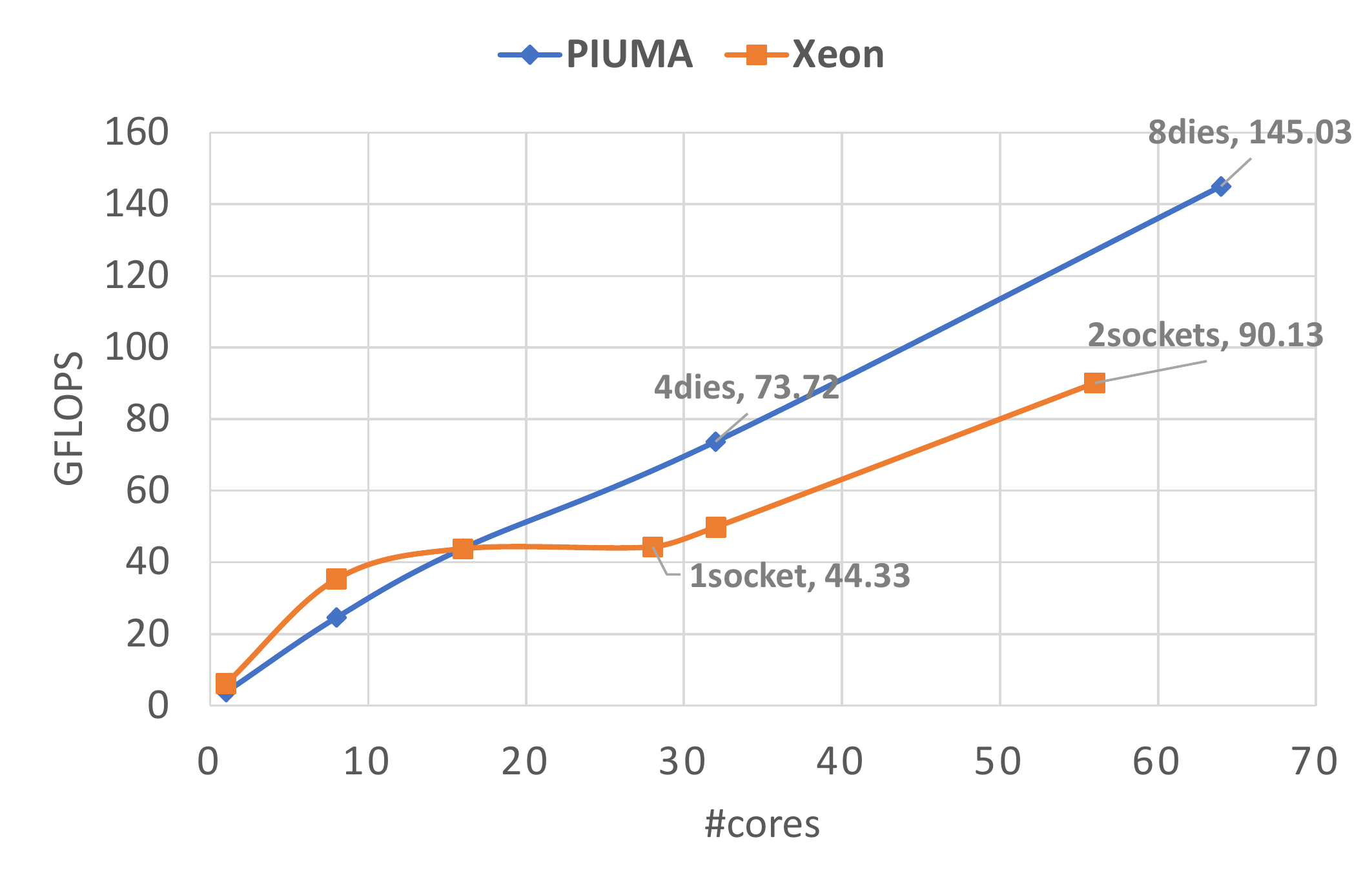}
	\caption{Comparison between PIUMA vs. Xeon.}
	\label{fig:PIUMAvsXeon}
\end{figure}

Figure~\ref{fig:PIUMAvsXeon} shows the strong scaling of both platforms for L=32, I=4 and fp64 data type. PIUMA appears to strong scale better than Xeon as shown by the data up to \(64\) cores. Although Xeon's architectural flops-per-byte ratio is 17.1 and PIUMA's effective flops-per-byte ratio is 1.25 (when only MTCs are used), and SU3\_Bench's arithmetic intensity is 0.675 for fp64, PIUMA wins here.

One interesting point to note here is that a PIUMA die consists of \(8\) PIUMA cores and therefore, 64 cores of PIUMA means these cores are distributed across 8 dies. However, we do not see any NUMA impact on PIUMA as we have seen on CLX. The PIUMA code also appears to be insensitive to the presence of a default constructor since the memory allocation happens differently on PIUMA than on Xeon. Since PIUMA is a latency and bandwidth optimized architecture, it does not suffer from those first touch and NUMA issues that we have faced on Xeon.

To summarize, what we learned from our exercise of porting SU3\_Bench to PIUMA is as follows: on PIUMA which is designed to target sparse workloads and hence has a low architectural flop-per-byte ratio and simple scalar pipelines, the performance of the SU3\_Bench kernel is limited by the effective instruction issue rate. PIUMA is bandwidth and latency optimized with similar network and memory bandwidth and follows a non-conventional stripped allocation policy by default. We learned that we do not need to worry about ``first touch'' for a memory allocation and NUMA issues on PIUMA that we had to for Xeon when using multiple sockets. With a few simple optimizations, SU3\_Bench kernel was able to obtain decent performance and exceed the performance of Xeon by \(1.5\times\) when using \(32\) cores on both systems.

%% file: Conclusion.tex
\section{Conclusion}
In this paper, we show the performance portability of SU3\_Bench on Xeon and Intel's Programmable Integrated Unified Memory Architecture (PIUMA). We show how the compiler provided default constructor touches memory at variable declaration time and the NUMA effect of the Xeon machine can impact performance and how simple change can improve the performance by \(2\times\) there. We also show how one can port the SU3\_Bench to PIUMA with a handful of changes to obtain a decent performance and how the effective instruction issue rate can impact performance on PIUMA. This paper can be considered as a use case study for porting SU3\_Bench kernel to PIUMA and we learned that despite PIUMA's low architectural flop-per-byte capacity and non-conventional memory allocation policy, SU3\_Bench is able to obtain close to peak performance in the system. Eventually, 32 PIUMA cores were able to overcome the performance of 32 cores of Xeon by \(1.5\times\).